\documentclass[10pt,aps,prb,twocolumn,noshowpacs]{revtex4-1}
\usepackage{amsmath,graphicx,amsbsy,amssymb,amsmath,epsfig,latexsym,wasysym,pifont,mathrsfs,natbib}
\usepackage{float} \newfloat{widefig}{thp}{lop} \usepackage{comment}
\usepackage{array, makecell} \usepackage{verbatim} \usepackage{datetime}
\usepackage{multirow,array} \usepackage{hyperref} \usepackage{color} \usepackage{setspace}
\usepackage{subcaption}
\usepackage{graphicx}
\usepackage{dirtytalk}
\usepackage[british]{babel}
\usepackage{csquotes}
\usepackage{comment}
\usepackage[export]{adjustbox}

\hypersetup{colorlinks,
  linkcolor=blue,%
  citecolor=blue,%
  urlcolor=blue}

\begin{document}

\title{\textbf{Improved  charge storage capacity of supercapacitor  electrodes by engineering surfaces: the case of Janus MXenes }}

\author{Mandira Das}
\affiliation{Department of Physics, Indian Institute of Technology
  Guwahati, Guwahati-781039, Assam, India.} 
\author{Subhradip Ghosh}
\email{subhra@iitg.ac.in} \affiliation{Dept. of Physics,
  Indian Institute of Technology Guwahati, Guwahati-781039, Assam,
    India.}   
\begin{abstract}
Surface Engineering in two-dimensional(2D) materials has turned out to be an useful technique to improve their functional properties. By designing Janus compounds MM$^{\prime}$C in MXene family of compounds M$_{2}$C where the two surfaces are constituted by two different transition metal M and M$^{\prime}$, we have explored their potentials as electrodes in a supercapacitor with acidic electrolyte. Using Density functional Theory (DFT) \cite{dft} in conjunction with classical solvation model we have made an in depth analysis of the electrochemical parameters of three Janus MXenes, passivated by oxygen - NbVC, MnVC and CrMnC. Comparisons with the corresponding end point MXenes Nb$_{2}$C, V$_{2}$C, Mn$_{2}$C and Cr$_{2}$C are also made. We find that the surface redox activity enhances due to formation of Janus, improving the charge storage capacities of MXene electrodes significantly. Our analysis reveals that the improved functionality has its root in the variations in the charge state of one of the constituents in the Janus compound which, in turn, has its origin in the electronic structure changes due to the surface manipulation. Our work, which is the first on the electrochemical properties of Janus MXenes for supercapacitor applications, suggests the surface engineering by forming appropriate Janus compounds as a possible route to extract high power density in a MXene electrode-acidic electrolyte based energy storage devices.
\end{abstract} 
\maketitle

\section{Introduction\label{intro}}
Supercapacitors have turned out to be a novel green energy storage devices as they combine orders of magnitude higher storage capacity than dielectric capacitors and superior power density than the traditional batteries. Electrode materials, one of the key components in an electrochemical supercapacitor, play a crucial role in their electrochemical performances. 
MXenes, the class of two-dimensional (2D) materials with chemical formula M$_{n+1}$X$_{n}$T$_2$ (M is a transition metal, X either Carbon or Nitrogen, T a functional group like -O, -F, -OH passivating the surface) discovered only about a decade ago, has grabbed the attention of materials research community. This is due to the superior electrochemical performance of the first discovered MXene Ti$_{3}$C$_{2}$T \cite{naguib2011two}over other established 2D materials like Graphene\cite{novoselov2004electric,ambrosi2014electrochemistry,zhang2010graphene,achee2018high}, transition metal di-chalcogenides\cite{ali20222d}(VSe$_2$\cite{li4485056lithium}, WS$_2$\cite{sharma2020large}), and transition metal oxides(RuO$_2$\cite{guduru2021electrochemical,ferris20153d}, MnO$_2$\cite{corpuz2021manganese,kwon2021mno,julien2017nanostructured}, NiO\cite{wang2022review}). Flexibility in composition makes MXenes suitable for further exploration of enhanced functionalities by tweaking their compositions and structures. Subsequently, experiments have been performed  on synthesised new MXenes \cite{yadav2023structural,tan2022v,yang2022heterogeneous,zhang2021nb2ct,zhu2019synthesis,bayhan2023laser}, heterostructured \cite{raj2022heterostructured,feng2023interface,ka2022hierarchical,zhang2022ti3c2tx,zheng2021multifunctional} and doped \cite{wen2017nitrogen,yang2022synthesis,li2018improved}Ti$_{3}$C$_{2}$T$_2$, solid solutions \cite{wang2021adjustable}, and ordered phases with varieties of compositions \cite{anasori2015two,el2021enhanced,ahmed2020mxenes}. Improved electrochemical performances of supercapacitors in most of these cases, therefore, call for further investigations into the electrochemical properties of MXene based supercapacitors by manipulating the compositions. 

One way to explore structure-property relations in 2D materials is to design compounds with asymmetric surfaces and study its impact on physical properties. To this end, quasi 2D Janus materials MoSSe\cite{jang2022growth} and  WSSe\cite{harris2023real} from the transition-metal dichalcogenide family have been synthesised. Experiments discovered that these compounds exhibit photo-voltaic properties\cite{zheng2021excitonic}, superior catalytic activity for hydrogen evolution reaction (HER) \cite{dong2017large},vertical piezoelectricity\cite{cai2019tribo}, and Rashba effect\cite{yao2017manipulation,hu2018intrinsic,chen2020tunable,cheng2013spin}, among other novel features. These clearly are the effects of the structural modifications of the pristine dichalcogenides. For example, the superior HER efficiency of MoSSe over MoS$_{2}$ is due to the presence of single S and Se vacancies that strengthen the binding of hydrogen with dichalcogenide monolayers, turning the inert surfaces highly active catalytically\cite{dong2017large}. As a charge storage device Janus MoSSe has shown greater potential too. Monolayer MoSSe can simultaneously possess good storage ability and fast Na/K ion diffusion and thus can be used as an anode material for Na/K-ion batteries observed in DFT calculations\cite{MoSSebatteries}. First principles calculations on MoSSe monolayer and bi-layer have shown that they can be used as anodes in Li ion batteries too  as the charge asymmetry on the two surfaces induce an internal electric dipole moment that generates an intrinsic electric field from Se layer towards S layer facilitating higher storage of Li ions \cite{MoSSeLibatteries}. Janus monolayer TiSSe and VSSe dichalcogenides too were predicted to be good anode materials for alkali ion metal batteries as computed by first-principles methods \cite{TiSSeVSSe}.

Though no Janus MXene has been synthesised, a plethora of first-principles based simulations is available \cite{physica2018,acsnano2019,cms2020} that establishes the thermal and mechanical stability of both MM$^{\prime}$XT$_2$ and M$_{2}$XTT$^{\prime}$ compounds up to significantly high temperatures. Calculations predict improved (over their end point  constituents M$_{2}$XT$_2$) thermoelectric properties\cite{wong2020high}, HER activity \cite{jpcc2020}, charge storage capacity in Li and Mg batteries \cite{siriwardane2021first}, and magnetic properties \cite{acsnano2019} for a number of them. However, no investigation of the capabilities of Janus MXenes as electrodes in supercapacitors is yet done.  In this communication, using DFT based methodologies and classical solvation model, we have investigated the charge storage capacities in two different groups of Janus MXenes MM$^{\prime}$CO$_{2}$; in one group M,M$^{\prime}$ are Ti,V and Nb while in the other group they are V,Cr and Mn. M$_{2}$CT$_{2}$ MXenes as well as solid solutions (M,M$^{\prime}$)$_{2}$CT$_{2}$ ( M,M$^{\prime}$ are Ti,V,Nb) and MAX phases (M,M$^{\prime}$)$_{2}$AC with M,M$^{\prime}$ being V,Cr,Mn, have been experimentally synthesised \cite{hong2020double,wang2021adjustable}. These indicate that there is substantial possibility of formation of Janus phases with these constituents. In absence of experimental reports, we have performed a detailed study on the possibilities of formation, thermal and mechanical stabilities of these MXenes. Our calculations show that out of the six -O functionalized Janus compounds considered, only three are mechanically and thermally stable. Electrochemical properties of these three in acidic electrolyte solution are then computed and compared with the results on end point MXenes. A better performance due to this surface engineering is observed in all three cases suggesting this as a possible route to enhance charge storage in MXene electrodes for supercapacitor applications. 
\section{Methodology}
\subsection{Structural Model of functionalised Janus MXenes\label{model-structure}}
\begin{figure}
    \centering
    \vspace{0.5 cm}
    \includegraphics[width=0.8\linewidth]{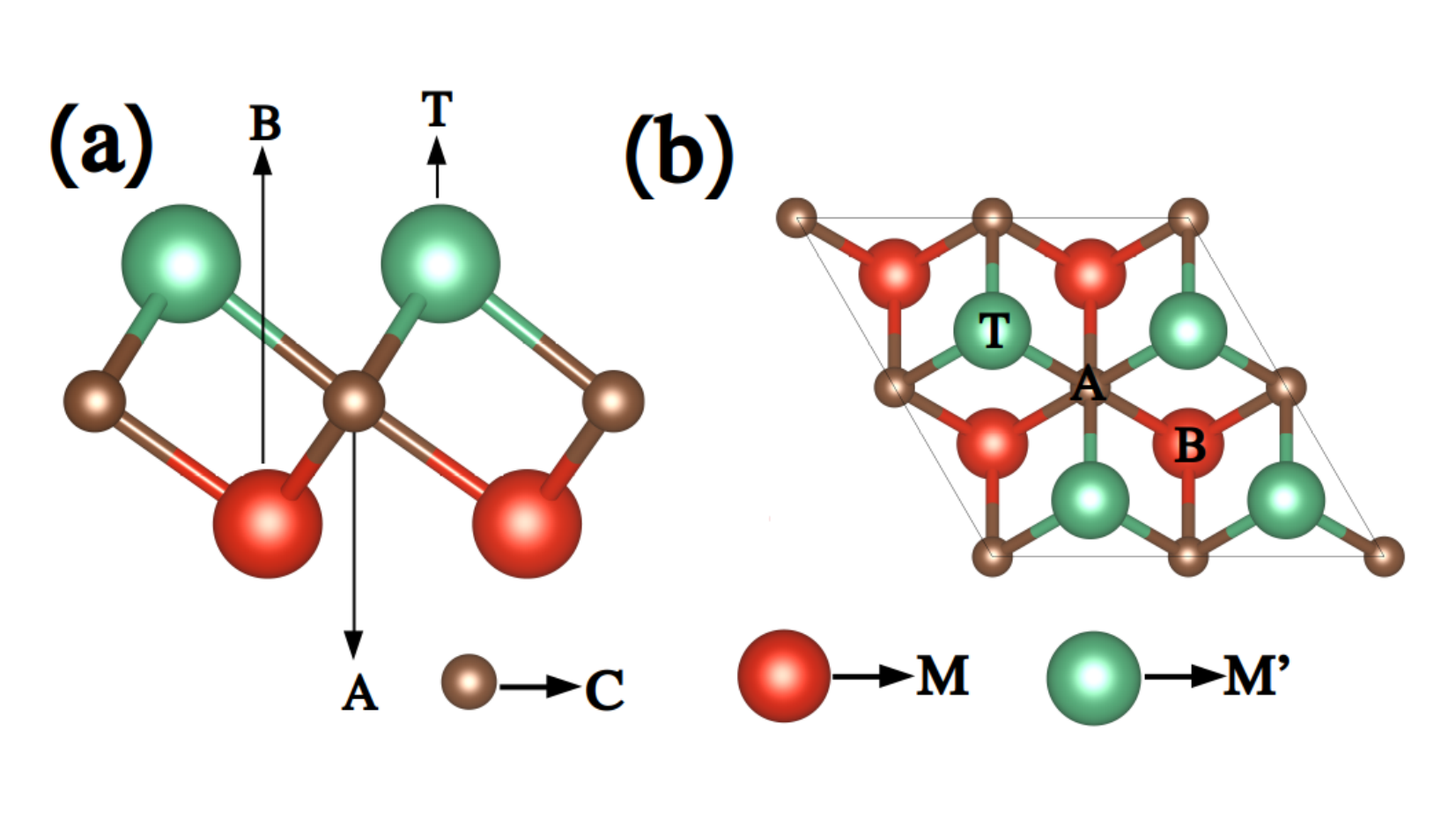}
    \caption{(a) Side view and (b) Top view of MM$^{\prime}$C Janus MXene structure. Brown, Red, and Green balls stand for the C, M, and M' elements, respectively.}
    \label{FIG:1}
\end{figure}

Monolayer of Janus MXene MM$^{\prime}$C is obtained from monolayers of M$_{2}$C pristine MXenes by replacing transition metal M atoms of one of the two surfaces with another transition metal M$^{\prime}$. The structure of MM$^{\prime}$C MXene is shown in Figure \ref{FIG:1}. 
 During the synthesis process, MXene surfaces are passivated by functional groups T, the most common among them are -F, -O, and -OH. In this work we have considered Janus MXenes with surfaces passivated by -O. In Figure \ref{FIG:1} we show different positions available to -O for a given surface : (1) site T, the position right over the transition metal, (2) site A,  the hollow site of a carbon atom and (3) site B, the hollow position corresponding to the transition metal element on the other surface (Figure \ref{FIG:1}(a)). Two asymmetric surfaces and three possible sites result in 3$^{2}$,$i.e$, nine models for a MM$^{\prime}$CO$_{2}$ Janus MXene. We first compute total energies of all 9 structures to obtain the ground state. We then check the dynamical and thermal stability of the ground state structure by computing the phonon spectra and by performing $Ab$ $initio$ molecular dynamics (AIMD) simulations. 
 
The total energies and the electronic structures  are calculated by the DFT based projector augmented wave method\cite{mortensen2005real} as implemented in Vienna ab-initio Simulation Package(VASP)\cite{kresse1999ultrasoft}. The exchange-correlation part of the Hamiltonian is described by Perdew-Burke-Ernzehof (PBE) Generalized Gradient Approximation (GGA)\cite{perdew1996generalized,paw}. We use a kinetic energy cut-off of 520 eV and a Monkhrost-pack\cite{mp} grid of 12$\times$12$\times$1 for self-consistent calculations. A larger k-mesh of 24$\times$24$\times$1 is used for the  Densities of States calculations. The convergence criteria for energy and force criteria were set to 10$^{-6}$ eV and 10$^{-5}$ eV/{\AA}, respectively. The second-order interatomic force constants(IFCs) are calculated by the density functional perturbation theory (DFPT)\cite{baroni2001phonons} method implemented in VASP using a 3$\times$3$\times$1 supercell and 6$\times$6$\times$1 k-mesh. The phonon spectra are generated using these IFCs by the finite-difference method as implemented in the PHONOPY \cite{togo2015first} package. $Ab$ $initio$ molecular dynamics (AIMD) simulations are performed with 3$\times$3$\times$1 supercell at 300K with Nose-Hoover thermostat\cite{nose1984unified,hoover1985canonical} as implemented in VASP for 10 ps with a time step of 2 fs.

\subsection{Modelling MXene electrode-acidic electrolyte interaction and calculation of the Capacitance\label{calculation-capacitance}}
Charge storage mechanism in a supercapacitor is two-fold : one, due to  the formation of a charged ion layer over the charged electrode upon application of a voltage, leading to electrochemical double layer capacitance or EDLC($C_{EDL}$), and two, due to  charge transfer between the charged electrode and accumulated ions across the layer, leading to redox capacitance($C_{redox}$). Parallel combination of these two result in the total electrical capacitance $C_{E}$.
\begin{equation}
    C_E = C_{EDL} + C_{redox}
    \label{Eqn:2}
\end{equation}
In our work, the Joint Density Functional Theory (JDFT) approach is used to simulate the solid electrode-liquid electrolyte interface. In this approach, the solid electrode and the liquid electrolyte are described by quantum DFT and the  classical DFT, respectively\cite{petrosyan2005joint,petrosyan2007joint}. For study of the charging mechanisms, the Janus MXene electrodes are considered negatively charged, and in contact with the acidic electrolyte H$_{2}$SO$_{4}$. Upon application of electrode potential ($\phi$), the  H$^+$ ions of the electrolyte start to accumulate over the electrode surface. The redox reaction for the Janus MXene MM$'$CO$_2$ can be described as \cite{zhan2018understanding}
\begin{multline}
    MM'CO_2^{q} + 2H^{+} \rightarrow MM'CO_{(2-2x)}(OH)_{(2-2x)}^{q'} \\
    + 2(1-x)H^{+}
    \label{Eqn:3}
\end{multline}
where $x$ is the surface H-coverage that varies between 0 and 1, and $q$ is the net surface charge on the electrode. The $x$ and $q$ vary with the applied electrode potential $\phi$ during charging. The free energy (G($x$,$\phi$)) of the charged electrode with coverage $x$ and applied electrode potential $\phi$ is given as.
\begin{multline}
     G(x,\phi) = E(x) + xE_{ZPE} + Q(V(x,\phi))\phi +\\ E_{EDL}(V(x,\phi)) + (1-x)\mu_H^{+}
     \label{Eqn:4}
\end{multline}
$E(x)$ is the total energy of the solvated electrode with H-coverage of $x$ in zero surface charge. $E_{ZPE}$ is the zero point energy difference of the electrode between $x=0$ and $x=1$ configurations. The electrical work to move charge $Q$ (net charge on electrode) from zero potential (bulk electrolyte) to the electrode with potential $\phi$ is given by the term $Q\phi$. $E_{EDL}$ is the energy of the EDL induced by the electrode charge $Q$. $\mu_H^{+}$ is the chemical potential of the solvated proton in the electrolyte.   $V(x,\phi)$, $Q(V(x,\phi))$, $E_{EDL}(V(x,\phi))$ and $\mu_H^{+}$ are expressed as 
\begin{equation}
    V(x,\phi) = \phi - \phi_{PZC}
    \label{Eqn:5}
\end{equation}
\begin{equation}
    Q(x,\phi) = \int_{V=0}^{V=\phi-\phi_{PZC}(x)} C_{EDL}dV
    \label{Eqn:6}
\end{equation}
Constant $C_{EDL}$ approximation based on the Debye - Hückel model, estimated to be 32 $\mu F/cm^2$ for 1 mole H$_2$SO$_4$ is used here.
\begin{equation}
    E_{EDL}(x,\phi) = \int_{V=0}^{V=\phi-\phi_{PZC}(x)} Q(x,\phi)dV
    \label{Eqn:7}
\end{equation}
\begin{equation}
    \mu_H^{+} = \frac{1}{2}G[H_2] + e\phi_{SHE} - 0.059\times pH
    \label{Eqn:8}
\end{equation}
$G[H_2]$ is given by equation (\ref{Eqn:9}). The quantities in that equation are estimated from DFT and the thermodynamic database.
\begin{equation}
    G[H_2] = E[H_2] + ZPE[H_2] + \frac{7}{2}k_BT - TS_{H_2}
    \label{Eqn:9}
\end{equation}
 $\phi_{SHE}$ is a standard hydrogen electrode (SHE) potential, with a value of 4.66 V, determined from the PZC calibration of CANDLE\cite{sundararaman2015charge} solvation in JDFTx\cite{sundararaman2017jdftx}. We choose to vary  $\phi$ from -1V to +1V with respect to $\phi_{SHE}$.  We use the JDFT method with an implicit solvation model as implemented in the simulation package JDFTx\cite{sundararaman2017jdftx} to obtain electronic structure and the potential at the PZC of each H coverage to calculate the free energy function. The charge-asymmetric non-locally determined local-electric (CANDLE) model describes the implicit electrolyte\cite{sundararaman2015charge}. The $E(x)$ and $\phi_{PZC}$ are determined from the JDFTx simulation for different $x$ coverage and configurations. After having the value of $\phi_{PZC}$, we determine the terms $V(x,\phi)$, $Q(V(x,\phi))$, $E_{EDL}(V(x,\phi))$ using equations (\ref{Eqn:5})-(\ref{Eqn:8}) and feed them along with $E(x)$ to equation (\ref{Eqn:4}). Once $G(x,\phi)$ is known, the partition function $Z(\phi)$ can be expressed as 
\begin{equation}
    Z(\phi) = \int_{x=0}^{x=1} exp(-\beta G(x,\phi))dx
    \label{Eqn:10}
\end{equation}
The average H-coverage $x_{avg}$ and net surface charge on the electrode $Q_{avg}(\phi)$ can then be expressed as a function of $\phi$,
\begin{equation}
    x_{avg}(\phi) = \frac{\int_{x=0}^{x=1}x\, exp(-\beta G(x,\phi))dx}{Z(\phi)}
    \label{Eqn:11}
\end{equation}
\begin{equation}
    Q_{avg}(\phi) = \frac{\int_{x=0}^{x=1}Q(x,\phi)\, exp(-\beta G(x,\phi))dx}{Z(\phi)}
    \label{Eqn:12}
\end{equation}
The slopes of $x_{avg}(\phi)$ and $Q_{avg}(\phi)$ with electrode potential $\phi$ give  $C_{redox}$ and $C_{EDL}$, respectively. Total $C_{E}$ can then be calculated using Equation (\ref{Eqn:2}) for a range of applied electrode potential $\phi$. For JDFT calculation, Generalised Gradient Approximation with the Perdew-Burke-Enzerhof functional (GGA-PBE)\cite{perdew1996generalized} is considered as an approximation to  exchange-correlation part of the Hamiltoniana. Ultrasoft pseudopotentials\cite{garrity2014pseudopotentials} are used to describe the ion-electron interaction. Kinetic energy cutoff of 20 Hartree and 30 Hartree are used for structure optimization and single point energy calculation at the optimized geometry with convergence criteria of 10$^{-6}$ Hartree. For all calculations, 3$\times$3$\times$1 supercells of the unit cells are considered.

\section{Results and Discussions}
\subsection{Structural stability and parameters\label{model-result}}
As mentioned in section \ref{intro}, we have worked with six different -O functionalised Janus MXenes :  TiNbC, VTiC, NbVC, CrMnC, MnVC, and CrVC. As a first step towards investigating their structural stability, we have performed total energy calculations for each one of the nine models discussed in section \ref{model-structure}. Out of the nine possible combinations of the two O atoms, ones involving the position \textquote{T} (Figure \ref{FIG:1}(a) and (b)) deform the structures completely. Therefore, those possibilities are discarded .The total energies for the rest four models are shown in Table \ref{TAB:1} for each one of the compounds. For each system, the model with  minimum energy is considered as the reference to scale energies in other models.We find that while both O atoms occupying B sites minimise the energy for NbVCO$_{2}$ and VTiCO$_{2}$, in case of CrMnCO$_{2}$, both O minimise the total energy by occupying A positions. For the other three compounds TiNbCO$_{2}$, CrVCO$_{2}$ and MnVCO$_{2}$, the O atom associated with the surfaces of Ti(Nb),Cr(V) and Mn(V) occupies A(B) positions to minimise total energies. It is to be noted that the structures of VTiCO$_{2}$ in AA,AB models and those of CrVCO$_{2}$ and MnVCO$_{2}$ in BA models, distort. Those models are, therefore, discarded from further consideration. 

\begin{figure}[ht!]
    \centering
    \includegraphics[width=1.0\linewidth]{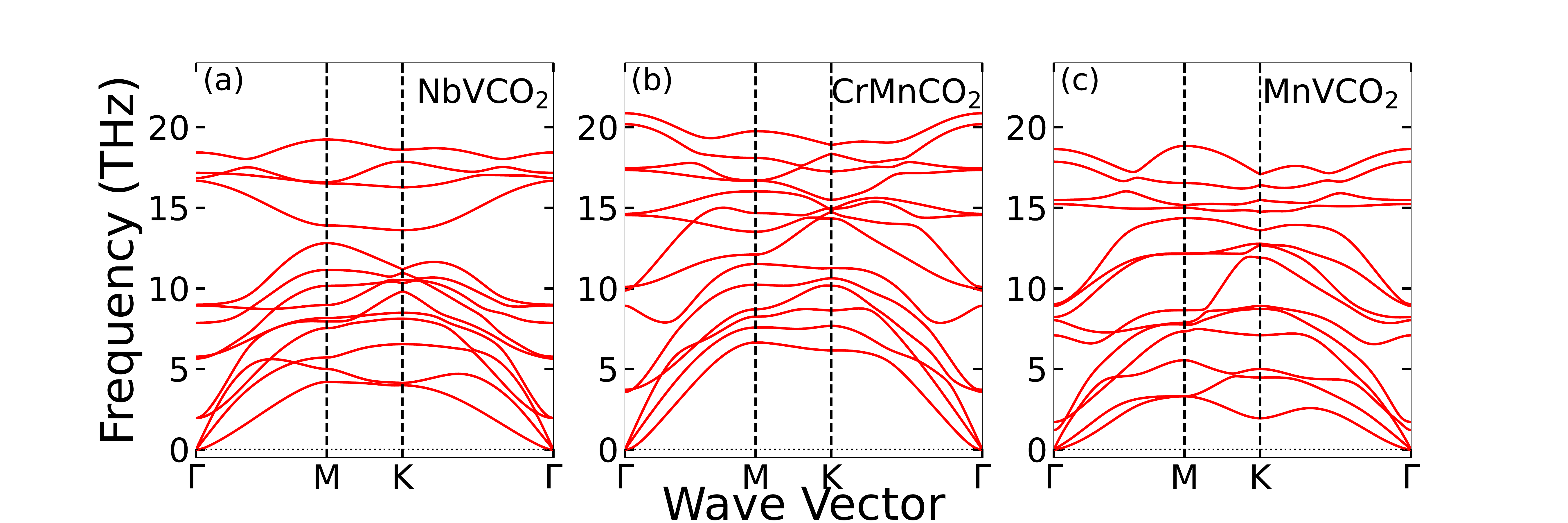}
    \caption{Phonon spectra of (a)NbVCO$_2$, (b)CrMnCO$_2$ and (c) MnVCO$_2$. Calculations are done with models AB for CrMnCO$_{2}$ and MnVCO$_{2}$. Model AA is used to compute phonon spectra of NbVCO$_{2}$.}
    \label{FIG:2}
\end{figure}

\begin{figure}[ht!]
    \centering
    \includegraphics[width=1.0\linewidth]{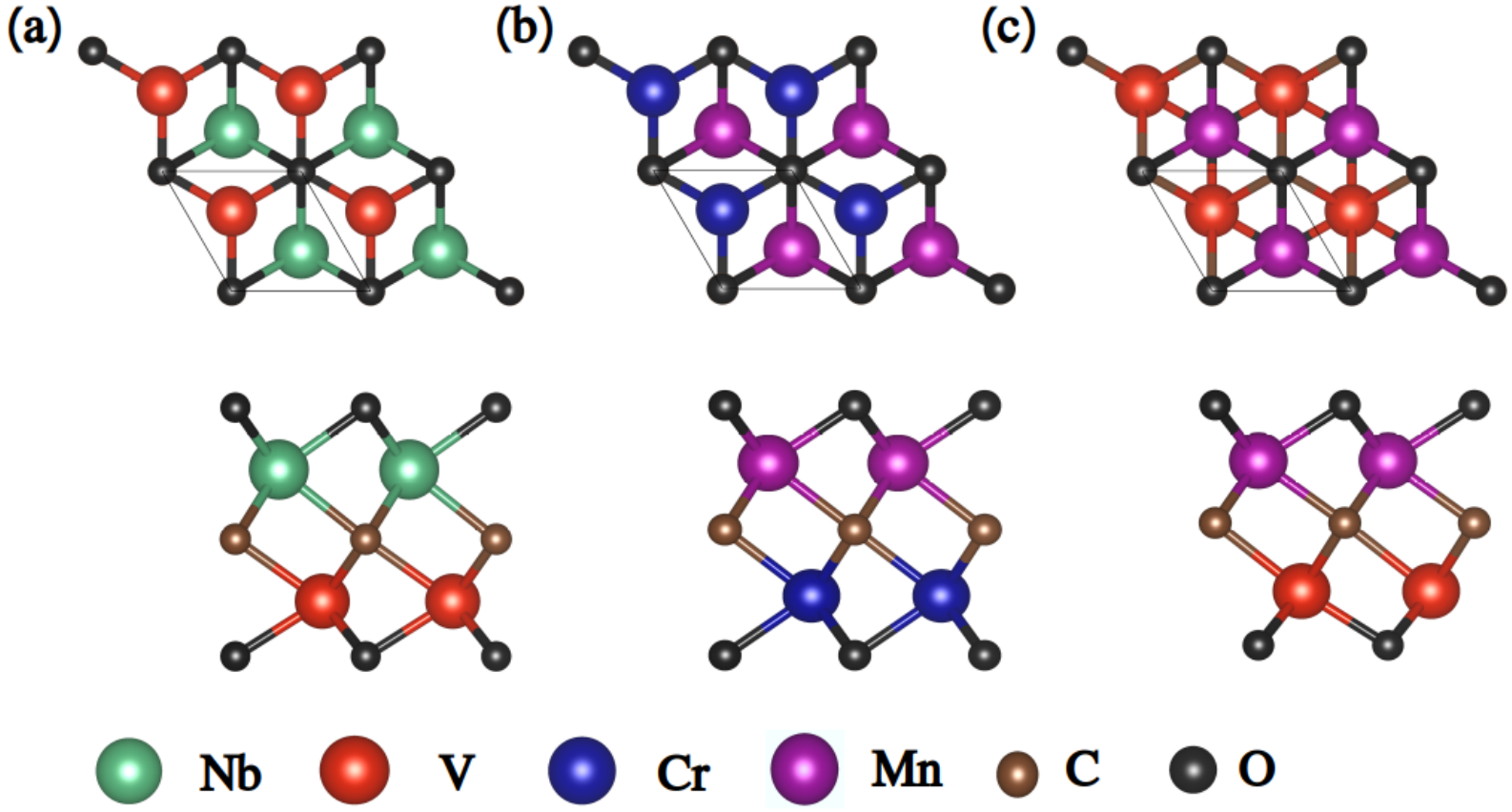}
    \caption{Dynamically stable lowest energy structures of (a)NbVCO$_2$, (b)CrMnCO$_2$, and (c)MnVCO$_2$. Both top (upper row) and side (lower row) views are shown.}
    \label{FIG:3}
\end{figure}

\begin{figure*}[ht!]
    \centering
    \begin{subfigure}[b]{0.3\textwidth}
    \includegraphics[width=\textwidth]{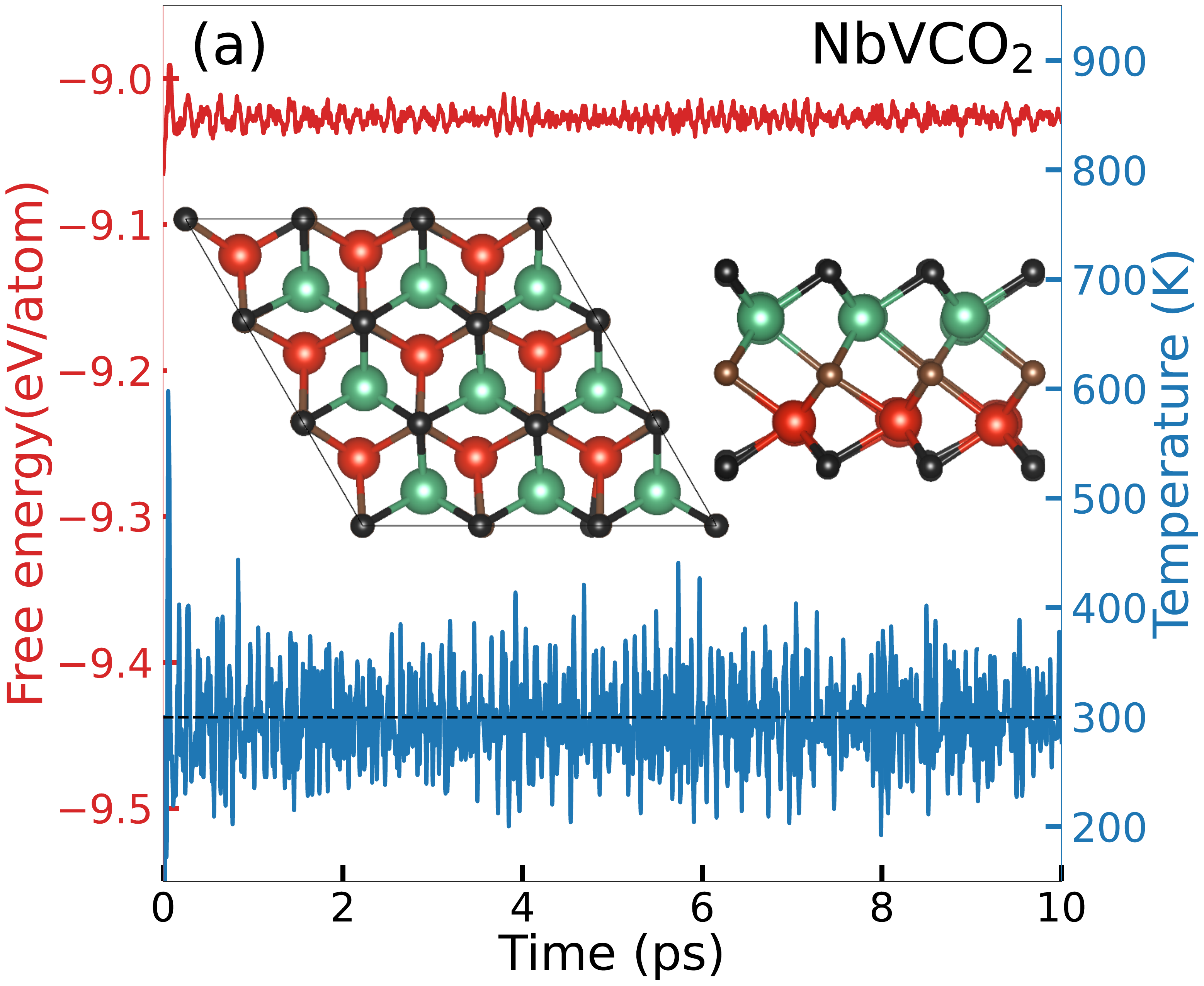}
    \end{subfigure}
    \hspace{-0.05cm}
    \begin{subfigure}[b]{0.3\textwidth}
    \includegraphics[width=\textwidth]{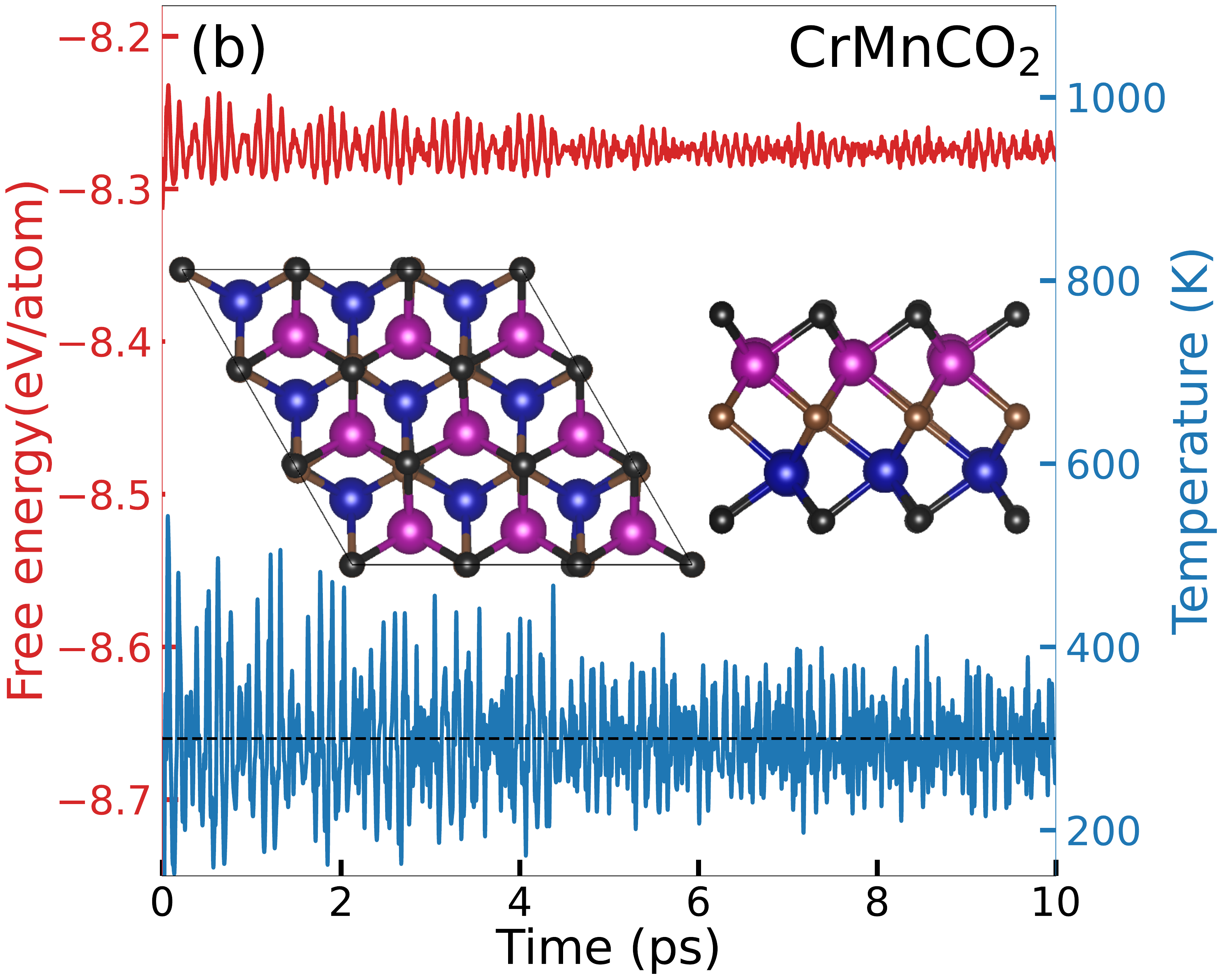} 
    \end{subfigure}
    \hspace{-0.05cm}
    \begin{subfigure}[b]{0.3\textwidth}
    \includegraphics[width=\textwidth]{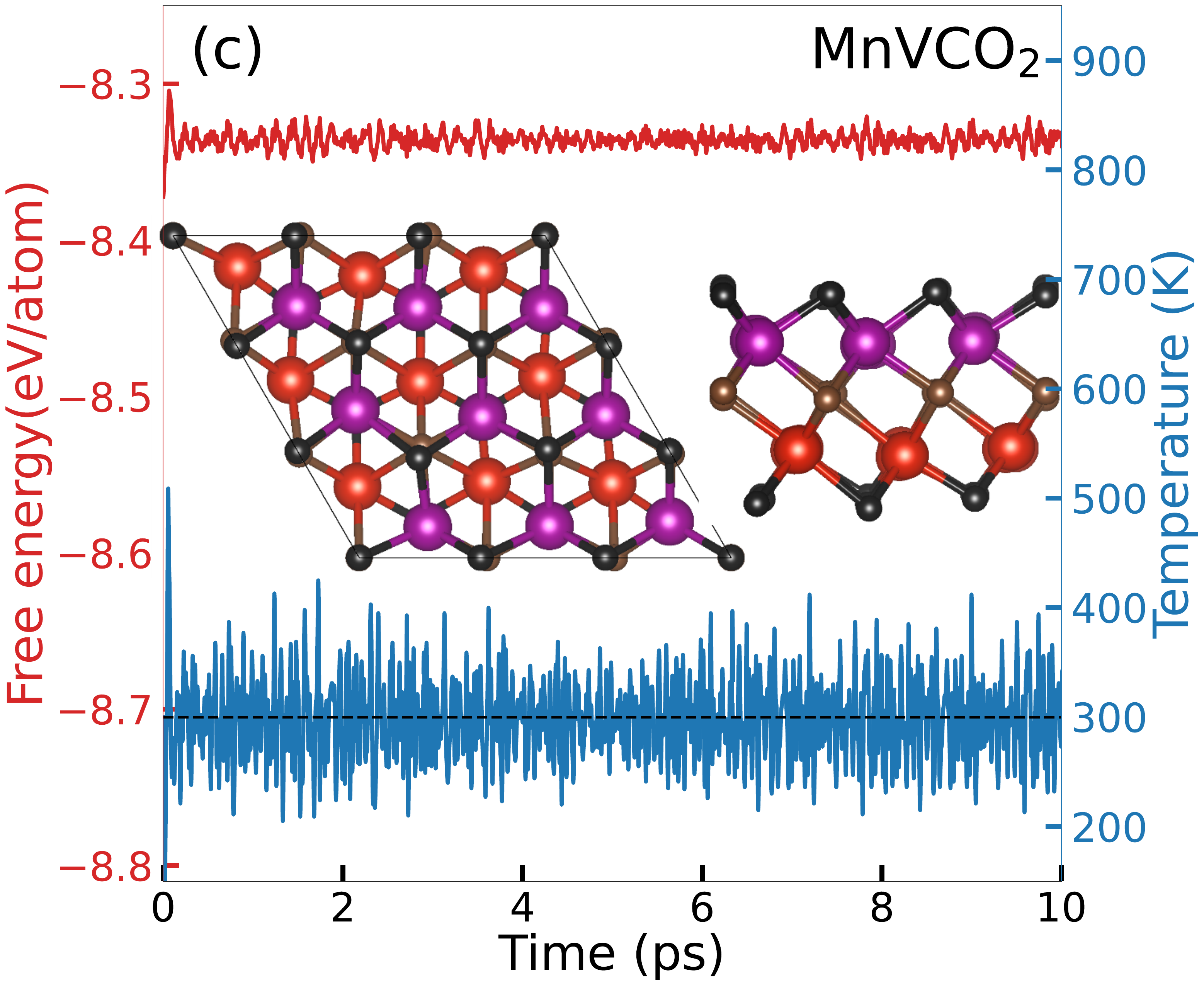}
    \end{subfigure}
    \caption{Variations in Free Energy(eV/atom) and Temperature(K) with time step(in ps) in AIMD calculations for (a)NbVCO$_2$, (b)CrMnCO$_2$, and (c)MnVCO$_2$. Insets show the top(left) and side(right) views of the final structures.}
    \label{Fig:4}
\end{figure*}

In the next step, we perform phonon calculations on all the remaining configurations where the structures are undistorted. We find that only three systems, NbVCO$_{2}$, CrMnCO$_{2}$ and MnVCO$_{2}$ are dynamically stable (Figure \ref{FIG:2}). While CrMnCO$_{2}$ and MnVCO$_{2}$ are dynamically stable in their respective ground states (Table \ref{TAB:1}), dynamical stability in NbVCO$_{2}$ is obtained in AA  model which is energetically only about 5 meV per atom higher than the ground state. The phonon spectra presented in Figure \ref{FIG:2}(a), is, therefore, calculated with the AA model. Since the energy of AA model is very close to that of the ground state, formation of this phase at low temperatures is possible and thus assessment of dynamical stability of NbVCO$_{2}$ by considering this configuration is justified. The phonon spectra of the dynamically unstable configurations are shown in Figures S1-S4, supplementary information. The dynamically stable structures of the three functionalised Janus MXenes are shown in Figure \ref{FIG:3}(a)-(c). We further assess the stability of these three compounds against thermal excitations by performing AIMD at room temperature. The results are shown in Figure \ref{Fig:4}(a)-(c). The variations in the Free energies and the temperatures over the observation time imply that the structures are stable. The final structures shown in the insets of the figures confirm that they are stable against thermal excitations. 
\begin{table}[H]
    \centering
    \setlength{\tabcolsep}{8pt} 
    \renewcommand{\arraystretch}{1.0} 
    \caption{Total energy of the Janus MXenes MM$^{\prime}$C for different models according to sites of O occupancy.}
    \begin{tabular}{|c@{\hspace{0.2cm}}| cccc@{\hspace{0.5cm}}|} 
    \hline
    \vspace{-0.33cm}
     \\System  &  \multicolumn{4}{|l|}{Total Energy (eV) for -O occupancy sites}\\
    \hline
           &   AA  &  BB &  AB  &  BA \\
    \hline
    TiNbCO$_2$  &  0.57  &   0.09   &    0.00      &   0.59    \\
    \hline
    NbVCO$_2$   &  0.03   &  0.00   &  0.26   &    0.13        \\
    \hline
    VTiCO$_2$    &  -  &  0.00   &   -   &   0.16        \\
    \hline
    CrMnCO$_2$   &  0.00  &  0.57  &  0.33  &   0.25   \\
    \hline
    CrVCO$_2$    &  0.04   &  0.44  &  0.00  &   -   \\
    \hline
    MnVCO$_2$    &   0.03  &  0.32  &  0.00  &   -  \\
    \hline
    \end{tabular}
    \label{TAB:1}
\end{table}

In Table \ref{TAB:2} information on ground states and structural parameters for the three Janus MXenes are shown. The information of the end point MXenes M$_{2}$CO$_{2}$ and M$^{\prime}_{2}$CO$_{2}$ are also shown for comparison. Due to presence of magnetic constituents Mn and Cr in MnVCO$_{2}$ and CrMnCO$_{2}$, additional magnetic calculations were done for these systems to ascertain the ground state.  We find that though O atoms stabilise Mn$_{2}$CO$_{2}$ by occupying  \textquote{B} positions, in both MnVCO$_{2}$ and CrMnCO$_{2}$, the O atoms associated with the Mn surfaces prefer to occupy \textquote{A} positions. The O atoms on V and Cr surfaces in Janus compounds, however, occupy the same positions as those in the end point compounds V$_{2}$CO$_{2}$ and Cr$_{2}$CO$_{2}$. In case of NbVCO$_{2}$, the structure is stabilised when O atoms occupy \textquote{A} positions on both surfaces although both end point compounds have \textquote{B} as the preferred position of O atoms. Thus in MnVCO$_{2}$, the passivation on two different surfaces lead to different symmetry. While the functinalisation on the Mn surface gives rise to trigonal prismatic symemetry, that on the V surface  leads to octahedral symmetry. For the other two Janus compounds, the symmetry is trigonal prismatic irrespective of the surface passivation. The magnetic ground state in MnVCO$_{2}$ is ferromagnetic (FM), different from the antiferromagnetic (AFM) ground state in Mn$_{2}$CO$_{2}$. The electronic ground state, too, is different in MnVCO$_{2}$ as compared to Mn$_{2}$CO$_{2}$. The semiconducting ground state of Mn$_{2}$CO$_{2}$ transforms to a metallic one when MnVCO$_{2}$ Janus is formed. In case of CrMnCO$_{2}$ Janus too, the electronic ground state is metallic, same as that of Cr$_{2}$CO$_{2}$. However, unlike the non-magnetic ground state of Cr$_{2}$CO$_{2}$, we get an AFM ground state in CrMnCO$_{2}$. The points to note are that although the electronic ground states in both MnVCO$_{2}$ and CrMnCO$_{2}$ are the same as the end compound MXenes V$_{2}$CO$_{2}$ and Cr$_{2}$CO$_{2}$, the magnetism driven by the inclusion of Mn in one of the surfaces lead to different ground states. In CrMnCO$_{2}$, the magnetic ground state of Mn$_{2}$CO$_{2}$ is retained but in MnVCO$_{2}$, it changes as well. It appears that while the properties of V and Cr surfaces are driving the electronic ground states, the passivation on different surfaces leading to different crystal symmetries in MnVCO$_{2}$ is responsible for the magnetic ground state of this compound being different from that of Mn$_{2}$CO$_{2}$. In both Mn-based Janus, the Mn moments are nearly equal and are substantially reduced from that in Mn$_{2}$CO$_{2}$. This is expected as the Mn-Mn inter-layer exchange interactions are absent in Janus compounds. In case of NbVCO$_{2}$, both surfaces are passivated by O atoms that conform to the same symmetry. The Janus system is metallic and non-magnetic like their end compound MXenes  Nb$_{2}$CO$_{2}$ and V$_{2}$CO$_{2}$. The lattice constants of CrMnCO$_{2}$, MnVCO$_{2}$ and NbVCO$_{2}$ Janus compounds are almost same as that of Cr$_{2}$CO$_{2}$, Mn$_{2}$CO$_{2}$ and V$_{2}$CO$_{2}$ MXenes, respectively. In CrMnCO$_{2}$ and MnVCO$_{2}$, Mn-C and Mn-O bond lengths decrease around 2-2.5 $\%$ in comparison to those in Mn$_{2}$CO$_{2}$. The Cr-C bonds in CrMnCO$_{2}$ shorten by 1$\%$ only with respect to Cr$_{2}$CO$_{2}$. On the other hand V-C bond lengths in MnVCO$_{2}$ increase by 1$\%$ while V-O bonds shorten  by 2.5$\%$ in comparison to V$_{2}$CO$_{2}$.The Mn-C bond lengths in this compound decrease by 2$\%$ in comparison with Mn$_{2}$CO$_{2}$. The relative changes (with respect to quantities in the end point MXenes) of M-C, M$^{\prime}$-C, M-O and M$^{\prime}$-O bond lengths are more significant in NbVCO$_{2}$ Janus. While there is hardly any change in Nb-C and Nb-O bonds, V-O bonds increase by 3$\%$ as compared to that in V$_{2}$CO$_{2}$. We can infer that overall there is no significant changes in the structural parameters when Janus MXene is constructed out of regular MXenes. 

\begin{table*}[]
    \centering
    \setlength{\tabcolsep}{8pt} 
    \renewcommand{\arraystretch}{1.0} 
    \caption{Sites of -O functionalisation, lattice-parameter(a), lengths of M-C(d$_{M-C}$),M$^{\prime}$-C(d$_{M^{\prime}-C}$), M-O(d$_{M-O}$), M$^{\prime}$-O (d$_{M^{\prime}-O}$) bonds, electronic and magnetic ground states (GS) of all the MM$^{\prime}$CO$_{2}$, M$_{2}$CO$_{2}$ and M$^{\prime}$CO$_{2}$ MXenes are shown. Magnetic moments of transition metal constituents are also shown for magnetic systems.}
    \begin{tabular}{|c@{\hspace{0.2cm}} |ccccccc@{\hspace{0.2cm}}|} 
    \hline
    \vspace{-0.33cm}
     \\Properties  &  \multicolumn{7}{|c|}{Systems}\\
     \hline
                   &  Nb$_2$CO$_2$  &  NbVCO$_2$   &  V$_2$CO$_2$  & MnVCO$_2$  &  Mn$_2$CO$_2$  & CrMnCO$_2$ & Cr$_2$CO$_2$ \\ 
     \hline
    Sites of         &  BB            &  AA          &   BB          &  AB           &    BB           &  AA       &    AA  \\
    -O functionalisation & & & & & & & \\
     \hline
     a({\AA})        &  3.12  &  2.95  &  2.91  &  2.87  &  2.88  &   2.68  &  2.69 \\
     \hline
     d$_{M-C}$ (\AA) &  2.19  &   2.17  &   2.06 &   2.00  &  2.04  &   2.00 &   2.02 \\
     d$_{M^{\prime}-C}$ (\AA)            &     -        &   2.06  &      -     &    2.08  &      -      &   2.00 &
           -       \\
    \hline       
     d$_{M-O}$ (\AA) &   2.09  &   2.08  &   1.95 &  1.93 &   1.93 &   1.92 &  1.92 \\
       d$_{M^{\prime}-O}$ (\AA)           &   -         &    2.01  &   -       &    1.92 &   -       &   1.90 &  -         \\
     
     \hline
     Magnetic GS  &  NM  &  NM  &  NM  &   FM   &   AFM  &  AFM  &  NM \\
     \hline
     Electronic GS &  metallic &  metallic & metallic &  metallic &  semi-conductor & metallic & metallic \\
     \hline
     Magnetic moment $\mu_{i}$ of &  - & - & - & $\mu_{Mn}$=1.2 & $\mu_{Mn}$=$\pm$2.8  &  $\mu_{Cr}$=$\pm$0.1 & - \\
       magnetic species $i$ (in $\mu_B$)      &    &   &   & $\mu_{V}$ = 0.95 &            &   $\mu_{Mn}$=$\pm$1.0 & - \\
     \hline

    \end{tabular}
    \label{TAB:2}
\end{table*}

\subsection{Electrochemical Capacitances}
\begin{figure}
    \centering
    \vspace{0.5cm}
    \includegraphics[width=1.1\linewidth]{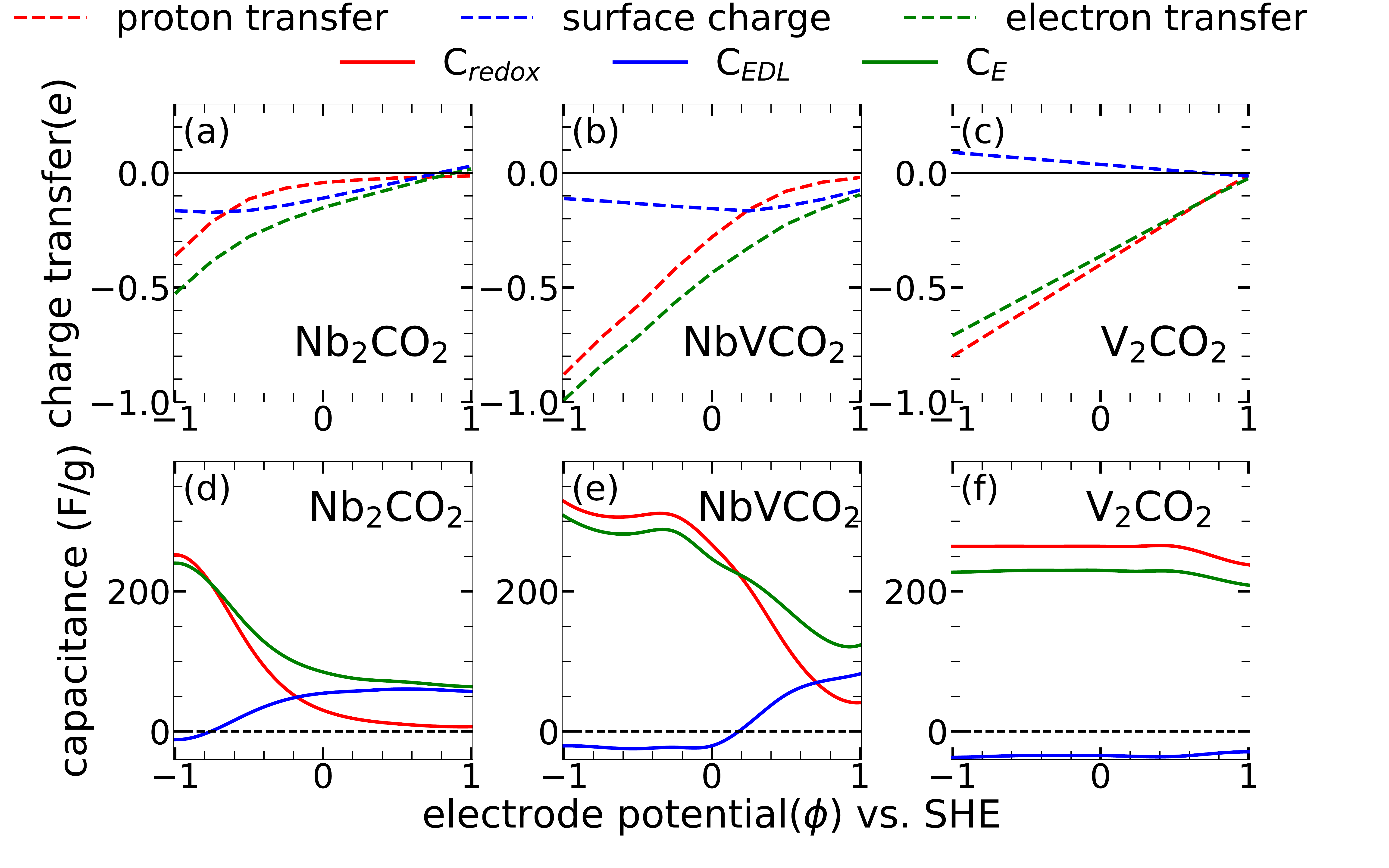}
    \caption{Various contributions to charge transfer and capacitance for Nb$_2$CO$_2$, NbVCO$_2$ and V$_2$CO$_2$}
    \label{FIG:5}
\end{figure}
\begin{figure}[ht!]
    \centering
    \includegraphics[width=1.0\linewidth]{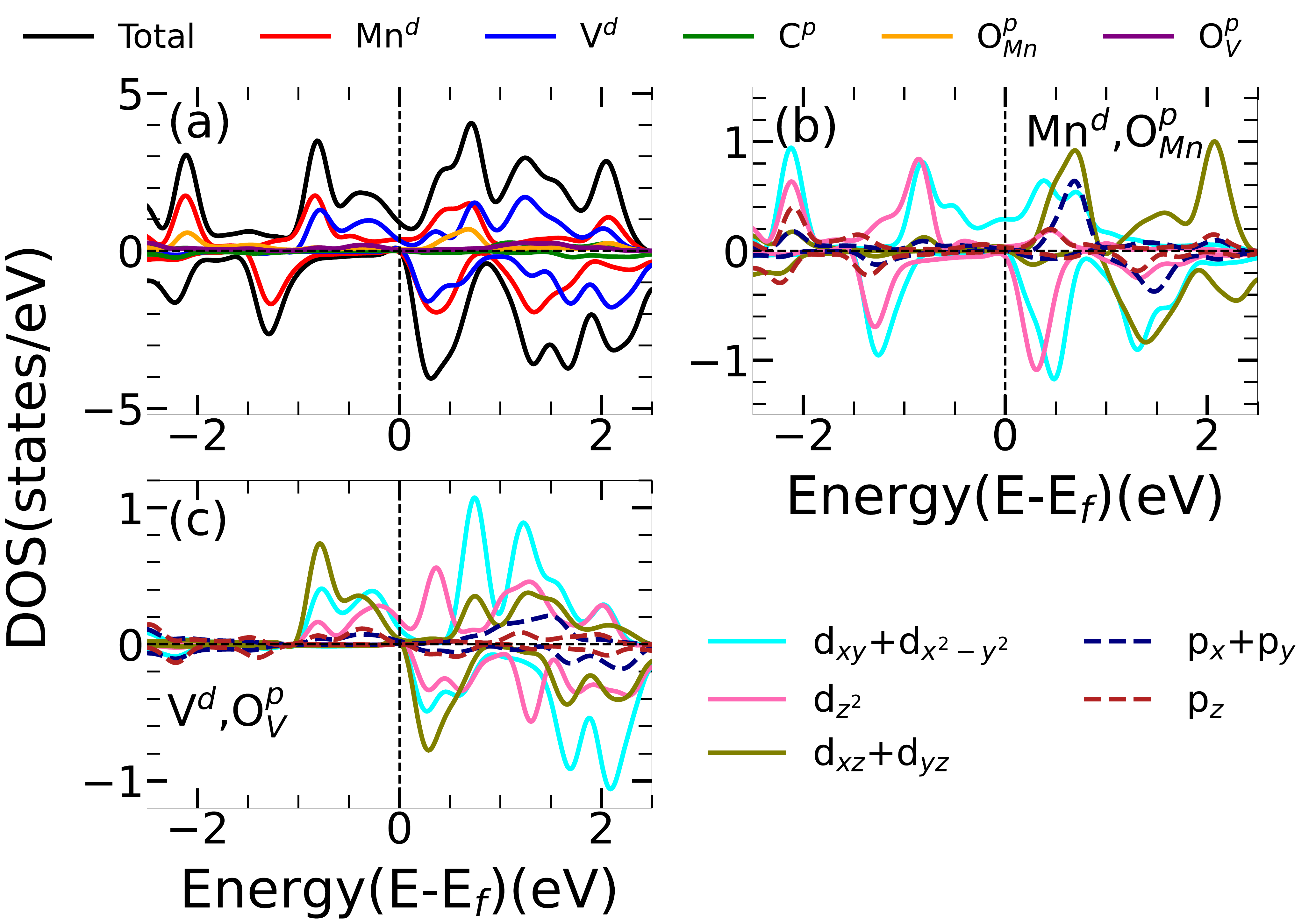}
    \caption{(a) Total and Partial densities of states of MnVCO$_2$ (b) d and p orbital projected densities of states of Mn and O(connected to Mn), (c) d and p orbital projected densities of states of V and O(connected to V)}
    \label{FIG:8}
\end{figure}

We now discuss the charge storage capacities of Janus compounds by making a comparison with corresponding end point MXenes. In an acidic electrolyte like H$_{2}$SO$_{4}$T the H$^{+}$ ions of the electrolyte accumulate over the negatively charged MXene electrode at the time of charging triggering electron transfer from the charged electrode to the ions. This is equivalent to proton transfer from electrolyte to the electrode. The amount of proton transfer is obtained from $x_{avg}$  computed by the Equation (\ref{Eqn:11}). The accumulated positive ions and the negatively charged MXene electrode act as two plates of a conventional capacitor and form an electrochemical double-layer (EDL) capacitor. A surface charge, $Q_{avg}$, computed by Equation (\ref{Eqn:12})  gets induced over the electrode due to the formation of EDL. The proton transfer number and induced surface charge vary with applied electrode potential $\phi$. The proton transfer and induced surface charge together produces the total charge storage capacity, denoted by the electron transfer number used in this paper. The gradient of proton transfer number, induced surface charge, and electron transfer number  with $\phi$ give C$_{redox}$, C$_{EDL}$, and C$_{E}$, respectively. In this work, we have calculated charge transfers and capacitances with applied electrode potential($\phi$) in the range -1V to +1V vs. SHE.
In Figure \ref{FIG:8}(a)-(f), variations in proton number, surface charge and electron number along with variations in  corresponding capacitances with $\phi$ are shown for NbVCO$_2$ Janus and the corresponding end point MXenes Nb$_2$CO$_2$ and V$_2$CO$_2$). To understand the results, we first discuss the charge storage mechanism of Nb$_2$CO$_2$ and V$_2$CO$_2$. As seen from Figure \ref{FIG:8}(a), the proton transfer number of Nb$_2$CO$_2$ decreases following a parabolic trajectory as voltage is increased from -1V, resulting in a decrease of C$_{redox}$ (Figure \ref{FIG:8}(d)). The change is substantial - from 251 F/g at -1V (vs. SHE) to 10F/g at 0.6V (vs. SHE), saturating thereafter. C$_{EDL}$ of Nb$_2$CO$_2$ shows the opposite trend. The surface charge increases with $\phi$ (Figure \ref{FIG:8}(a)), the change being slower than the proton transfer. Accordingly, C$_{EDL}$ increases continuously  and eventually becomes the dominant contributor to C$_{E}$ at $\phi > -0.25V$. Thus, both EDL and Redox mechanisms contribute to the charge storage capacity of this compound producing a maximum(minimum) C$_{E}$ of 240(64) F/g (Figure \ref{FIG:8}(d)). The storage mechanism of V$_2$CT$_x$, on the other hand, is significantly dominated by the Redox mechanism. The proton transfer number varies linearly with $\phi$, the slope of the curve being large. The variation in surface charge is linear too. However, the change is rather small with increasing $\phi$ (Figure \ref{FIG:8}(c)). As a result both C$_{redox}$ and C$_{EDL}$ remain constant at values 264 F/g and 36 F/g, respectively up to $\phi=0.4V$. After that their values decrease slightly to 237 F/g and 30 F/g, respectively at $\phi=1$V (Figure \ref{FIG:8}(f)), resulting in a maximum C$_{E}$ of 228 F/g. 

Both proton transfer and surface charge variations in case of NbVCO$_2$ follow two distinct patterns in two potential windows. The proton transfer number varies linearly up to between -1.0- 0.25V, following the pattern of V$_{2}$CO$_{2}$, albeit with a higher slope. Between 0.25V and 1.0V, the variation is like the one obtained for Nb$_{2}$CO$_{2}$. Same trend is found for variations in the surface charge. Consequently, in the negative (positive) potential window both C$_{redox}$ and C$_{EDL}$ behaviours resemble that of V$_{2}$CO$_{2}$(Nb$_{2}$CO$_{2}$). However, due to sharper variations in the proton transfer implying higher redox activity, as compared to the end point MXene compounds, C$_{redox}$ varies between 328 F/g and 307 F/g in the negative potential window. In the positive potential window, C$_{redox}$ decreases rapidly to about 50 F/g at 1.0V. C$_{EDL}$, on the other hand increases continuously to a maximum of 80 F/g at 1.0 V. Therefore, the total capacitance C$_{E}$ varies between 307-285 F/g (170-130 F/g) in the negative (positive) potential window. This implies that if one chooses the negative potential window, the enhanced redox activity will produce a significantly larger C$_{E}$ for this Janus compound in comparison to their M$_{2}$CO$_{2}$ counterparts. Moreover, this result suggests that more power density, in comparison with the end compound MXenes, can be extracted from Janus NbVCO$_{2}$ in the negative $\phi$ window. On the other hand, in the positive $\phi$ region, larger energy density along with substantial power density can be obtained by synthesising Janus out of Nb$_{2}$CO$_{2}$ and V$_{2}$CO$_{2}$ MXenes, making this Janus very useful as a supercapacitor electrode.  
\begin{figure}
    \centering
    \vspace{0.5cm}
    \includegraphics[width=1.0\linewidth]{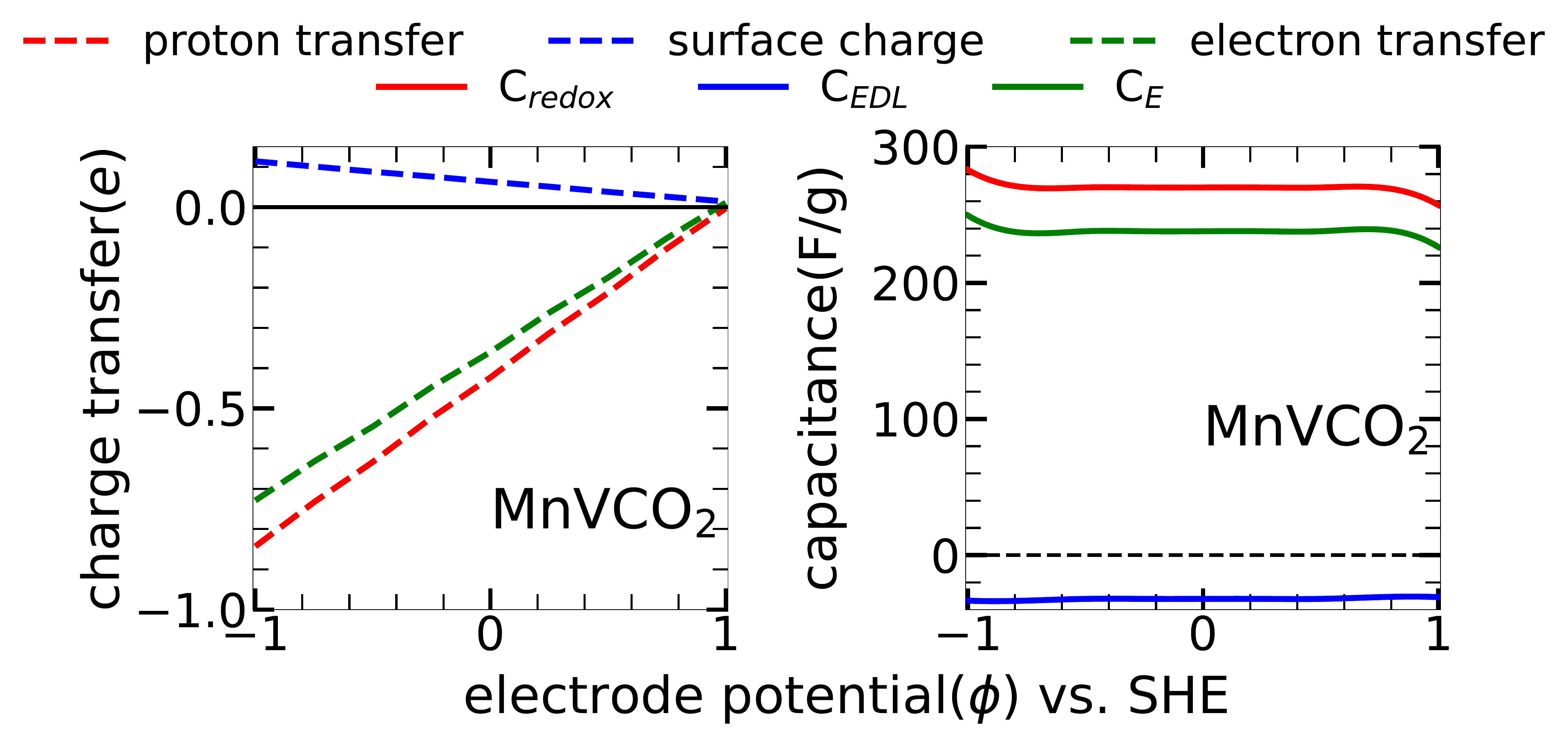}
    \caption{Various contributions to charge transfer and capacitance for MnVCO$_{2}$}
    \label{FIG:6}
\end{figure}

In Figure \ref{FIG:6}(a) and (b), we present the electrochemical behavior of MnVCO$_{2}$. Here we have not shown the electrochemical behaviour of the end point compound Mn$_{2}$CO$_{2}$. The electronic ground state of Mn$_{2}$CO$_{2}$ is semiconducting. Thus, it cannot be a high rate storage device like a metal. The variations in surface charge and proton transfer both will be much less than the metallic MXenes.  The variations in proton transfer number, surface charge and consequently the capacitances of MnVCO$_{2}$ Janus, expectedly, follow the same trend as  that of V$_{2}$CO$_2$. The sharper variations in the proton transfer, however, produces C$_{redox}$ larger than that obtained for V$_{2}$CO$_{2}$. As a result, the total capacitance C$_{E}$ varies between 249 F/g - 226 F/g over the entire potential window. 
\begin{figure}
    \centering
    \vspace{0.5cm}
    \includegraphics[width=1.0\linewidth]{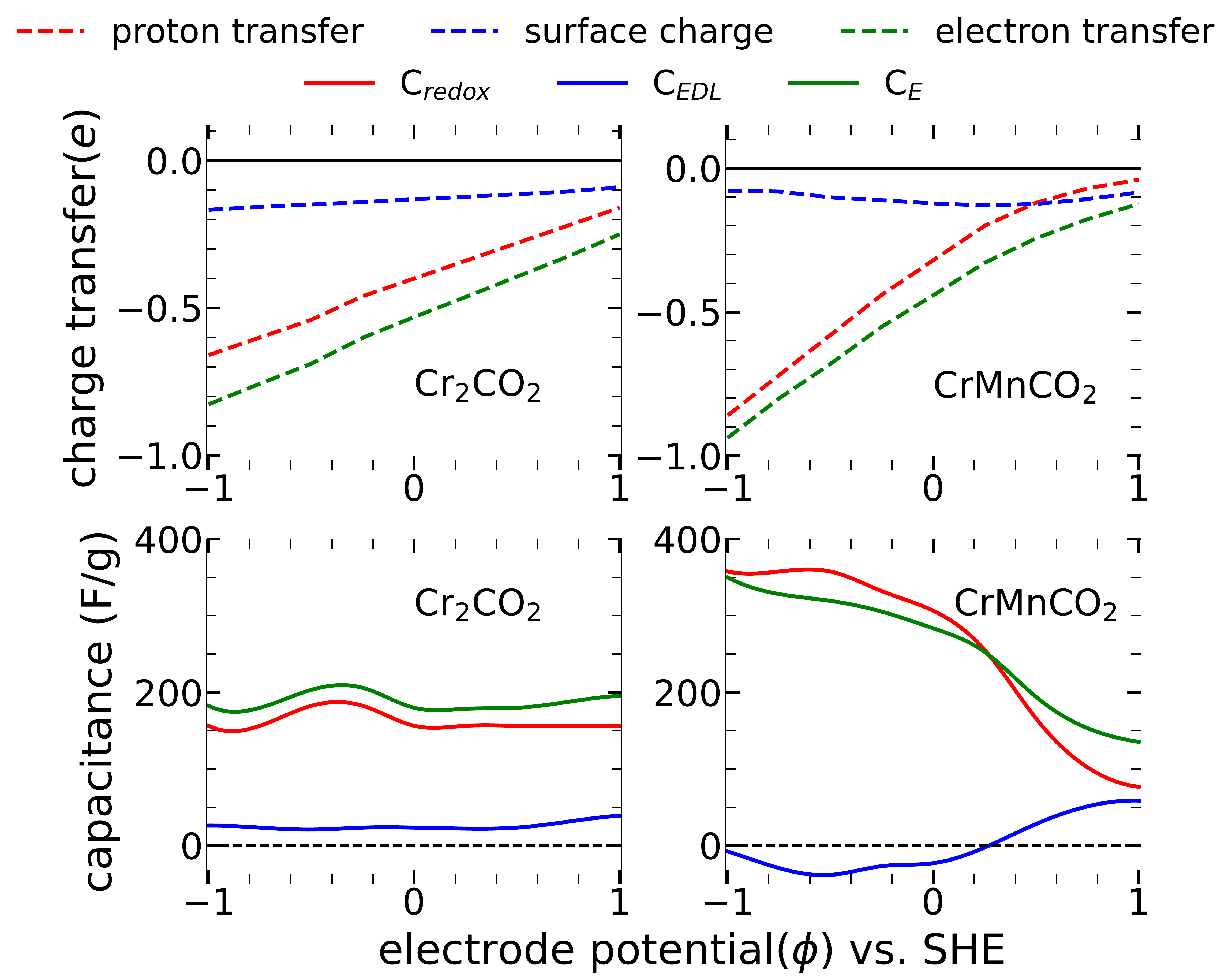}
    \caption{Various contributions to charge transfer and the capacitances for Cr$_2$CO$_2$ and CrMnCO$_2$}
    \label{FIG:7}
\end{figure}
Various contributions to the charge transfer and capacitance and their variations with $\phi$  for Cr$_2$CO$_2$ and CrMnCO$_2$ are shown in Figure \ref{FIG:7}. Like V$_{2}$CO$_{2}$, the charging mechanism in Cr$_{2}$CO$_{2}$ is dominated by the redox charge transfer. But unlike V$_{2}$CO$_{2}$, EDL has a co-operative effect to C$_{redox}$. Smaller variation in the proton transfer, in comparison to V$_{2}$CO$_{2}$, however, limits C$_{redox}$ to a maximum value of 175 F/g  for a small potential window and to a constant value of 156 F/g for the rest of the potential range. Consequently, C$_{E}$ varies between 179 -200 F/g. The variations of the electrochemical parameters in CrMnCO$_{2}$ Janus, contrary to the intuition, are quite different. The proton transfer number and the surface charge vary in such a way that between -1V-0.25V, the C$_{redox}$ after staying constant initially decrease; C$_{EDL}$ opposes C$_{redox}$ in this potential window. Between 0.25-1.0V, C$_{redox}$ decreases faster while C$_{EDL}$ co-operates with it. In this potential range, the EDL mechanism of charge storage catches up with the redox one gradually, having almost equal contributions at $\phi$ close to 1V. It is noteworthy that at $\phi$=-1V, EDL has no contribution to total capacitance and total capacitance is 350 F/g, the highest among the three Janus compounds considered in this work. Due to continuous weakening of the redox mechanism and thus gradual fall in C$_{redox}$, C$_{E}$ also decreases substantially to a minimum of 135 F/g at $\phi$=1V. 

The similarity among the three Janus compounds with regard to their electrochemical properties is the dominance of redox mechanism of charge storage along with significant enhancement in the capacitances with respect to the end point M$_{2}$CO$_{2}$ MXenes. The dissimilarity is in the trends in the variations in the capacitances, particularly C$_{redox}$ over the potential window considered. To understand the origin of these, we look at the variations in the charge states of the transition metal constituents in M$_{2}$CO$_{2}$ and the corresponding Janus MXene. In NbVCO$_{2}$, in the redox dominated potential window that is predominantly $\phi \leq 0$V, the charge state of Nb hardly changes (1.83-1.86) while charge state of V changes substantially from 1.51 to 1.63. This change is quantitatively slightly more than the change in V$_{2}$CO$_{2}$ (1.57-1.67). However, the change is more rapid in the Janus compound. For $\phi > 0$V, change in the charge state of V in NbVCO$_{2}$ is substantially smaller (1.63-1.69) than that in V$_{2}$CO$_{2}$ (1.67-1.78). This is the reason behind significant drop in the C$_{redox}$ for $\phi > 0$V in NbVCO$_{2}$. Therefore, the variations in C$_{redox}$ of NbVCO$_{2}$ can be understood in terms of the variations in the charge states of Nb and V. In Janus MnVCO$_{2}$,the charge of V does not undergo a significant deviation from V$_{2}$CO$_{2}$, either qualitatively or quantitatively. In the Janus, it varies between 1.54 and 1.70 over the entire potential window. In V$_{2}$CO$_{2}$, the variation is between 1.57 and 1.78. The Mn charge state, on the other hand, undergoes a variation that is larger and more rapid than that in end point MXene Mn$_{2}$CO$_{2}$. In Mn$_{2}$CO$_{2}$, the charge state of Mn varies between 1.51 and 1.63 over the entire potential window. In MnVCO$_{2}$, it varies between 0.97 and 1.28 over the same window. The larger values of C$_{redox}$ along with its nature of variation for MnVCO$_{2}$, thus, is due to the chemistry of the Mn surface. The case of CrMnCO$_{2}$ is intriguing. Since Mn$_{2}$CO$_{2}$ is a semiconductor and variations in the capacitances of MnVCO$_{2}$ followed the qualitative nature of V$_{2}$CO$_{2}$, it is only natural to expect that the same would happen in CrMnCO$_{2}$. The departure from the expectation has roots in the behaviour of the charge states of Mn and Cr. In Cr$_{2}$CO$_{2}$, the charge state of Cr varies by 0.04 as $\phi$ varies between -1V and 0.25V. The variations for the rest of the potential window is even smaller. This explains the maximum C$_{redox}$ around -0.5V (Figure \ref{FIG:7}))and a smaller value of it in comparison to V$_{2}$CO$_{2}$. Neither the charge state nor its variation undergoes any noticeable change in CrMnCO$_{2}$ except that beyond $\phi$=-0.25V, the charge state almost does not change. The Mn charge state in MnCrCO$_{2}$, on the other hand, behaves differently in different potential windows. The charge state of Mn varies more rapidly in CrMnCO$_{2}$ than in Mn$_{2}$CO$_{2}$. In the negative potential window, between $\phi$=-1.0V and -0.5V, Mn charge state changes by 0.19 in the Janus compound; the change is only 0.02 in Mn$_{2}$CO$_{2}$. However, in the Janus, the maximum change in the charge state of Mn for the rest of the potential window is 0.02. Such a large variation in Mn charge state for a small window explains the largest possible C$_{redox}$ obtained in the work. The stagnant charge state of both Mn and Cr for the larger part of the potential window explains why C$_{redox}$ continuously decreases in CrMnCO$_{2}$. 

\begin{figure}[ht!]
    \centering
    \includegraphics[width=1.0\linewidth]{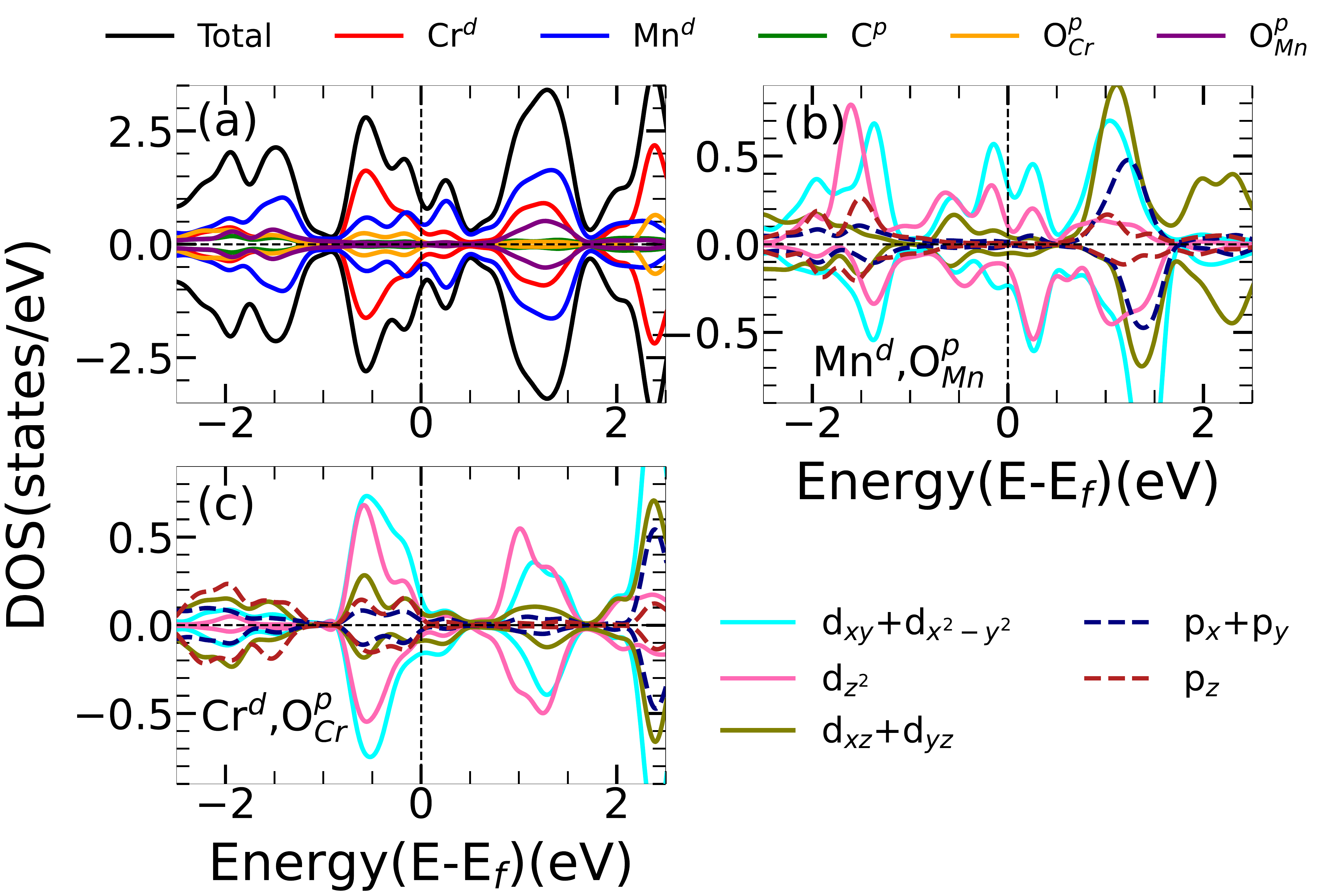}
    \caption{(a) Total and Partial densities of states of CrMnCO$_2$ (b) d and p orbital projected densities of states of Mn and O(connected to Mn), (c) d and p orbital projected densities of states of Cr and O(connected to Cr)}
    \label{FIG:9}
\end{figure} 

Thus, for both the Mn-based Janus, it is the Mn that takes pro-active role to enhance the redox mechanism and consequently the redox charge storage capacity, despite the absence of redox activity in Mn$_{2}$CO$_{2}$ MXene. The reason must lie in the changes of the electronic structure of Mn in both Mn-based Janus compounds. The significant re-distribution of Mn states in metallic Janus compounds in comparison with Mn states in semiconductor Mn$_{2}$CO$_{2}$ changes Mn charge states substantially. From Figures \ref{FIG:8} and \ref{FIG:9}, we can clearly see that the Mn $d_{xy}+d_{x^{2}-y^{2}}$ states are dominant near the Fermi level of both Janus compounds. Hybridisations with $p$ states of O and $d$ states of the other transition metal delocalise the Mn states to a large extent. This explains why even before application of a voltage, Mn charge state changes from 1.51 in Mn$_{2}$CO$_{2}$ to 1.28 in both Janus. It is to be noted that no other transition metal constituent has undergone such significant change in their charge states upon formation of Janus : V charge state changes from 1.78 in V$_{2}$CO$_{2}$ to 1.70 and 1.69 in MnVCO$_{2}$ and NbVCO$_{2}$, respectively, Nb state changes from 2.0 in Nb$_{2}$CO$_{2}$ to 1.88 in NbVCO$_{2}$, Cr state changes from 1.38 in Cr$_{2}$CO$_{2}$ to 1.28 in CrMnCO$_{2}$. The application of external voltage, therefore, affects the charge state of Mn more than that of others.    

\section{CONCLUSION}
This work is the first attempt to assess the potential of Janus MXenes as electrodes in supercapacitors. Since experimental synthesis of Janus MXenes is yet to happen, we chose systems in which either the Janus MAX phase or a MXene solid solution are synthesised. Careful test on their thermal and dynamical stabilities are performed to make sure the choice of Janus phase in these compounds are justified. Out of six compounds considered, only three were found to be thermodynamically stable upto room temperature. The key result obtained is that the charge storage capacity increases significantly in Janus MXenes MM$^{\prime}$CO$_{2}$ in comparison with the end point MXenes M$_{2}$CO$_{2}$ and M$^{\prime}$CO$_{2}$. In all cases the enhanced storage capacity is due to enhanced redox activity upon formation of Janus. The reason behind such enhancements lie in the rapid changes in the charge states of the transition metal constituents forming the Janus. The formation of Janus breaks the symmetry of M$_{2}$C MXenes and affects the electronic structures of the constituents substantially. This in turn affects the charge states. The degree of variation in the charge states decide the quantum of increment in redox capacitance. This work, therefore, throws reasonable amount of  light on the role of surface engineering and chemistry of materials in the context of energy storage device. Our work shows that this can be used as a viable route to obtain greater storage capacity when MXenes are used as electrodes in supercapacitors.  

%


\end{document}